\documentclass[journal,twoside]{IEEEtran}

\usepackage{amsmath}
\usepackage{amssymb}
\usepackage{bm}
\usepackage{graphicx}
\usepackage{cite}

\graphicspath{{fig/}}

\newcommand{\bw}{\ensuremath{\bm{w}}}

\newcommand{\bv}{\ensuremath{\bm{v}}}

\newcommand{\sV}{\ensuremath{\mathcal{V}}}

\newcommand{\bu}{\ensuremath{\bm{u}}}

\newcommand{\bvh}{\ensuremath{\hat{\bm{v}}}}

\newcommand{\bh}{\ensuremath{\bm{h}}}

\newcommand{\me}{\ensuremath{\mathrm{e}}}
\newcommand{\opt}{\ensuremath{\mathrm{opt}}}

\newcommand{\bhb}{\ensuremath{\bar{\bm{h}}}}
\newcommand{\diff}{\ensuremath{\mathrm{d}}}

\newtheorem{lemma}{Lemma}

\newtheorem{prop}{Proposition}
\newtheorem{corollary}{Corollary}

\begin{document}

\title{On Transmit Beamforming for MISO-OFDM Channels With Finite-Rate
  Feedback}

\author{Kritsada Mamat and Wiroonsak Santipach~\IEEEmembership{Member,~IEEE,}%
\thanks{This work was supported by the 2010
  Telecommunications Research and Industrial Development Institute
  (TRIDI) scholarship and joint funding from the Thailand Commission
  on Higher Education, Thailand Research Fund, and Kasetsart
  University under grant MRG5580236.}%
\thanks{The material in this paper was presented in part at the
  Electrical Engineering/Electronics, Computer, Telecommunications and
  Information Technology Conference (ECTI), Huahin, Thailand, May
  2012, and the IEEE International Conference on Communications (ICC),
  Budapest, Hungary, June 2013.}%
\thanks{The authors are with the Department of Electrical Engineering;
  Faculty of Engineering; Kasetsart University, Bangkok, 10900,
  Thailand (email: g5317500192@ku.ac.th; wiroonsak.s@ku.ac.th).}}

\markboth{IEEE Transactions on Communications}{Mamat and Santipach: 
On Transmit Beamforming for MISO-OFDM Channels With Finite-Rate
  Feedback}

\maketitle

\begin{abstract}
With finite-rate feedback, we propose two feedback methods for
transmit beamforming in a point-to-point MISO-OFDM channel. For the
first method, a receiver with perfect channel information, quantizes
and feeds back the optimal transmit beamforming vectors of a few
selected subcarriers, which are equally spaced. Based on those
quantized vectors, the transmitter applies either constant, linear, or
higher-order interpolation with the remaining beamforming
vectors. With constant interpolation, we derive the approximate sum
achievable rate and the optimal cluster size that maximizes the
approximate rate. For linear interpolation, we derive a closed-form
expression for the phase rotation by utilizing the correlation between
OFDM subcarriers. We also propose a higher-order interpolation that
requires more than two quantized vectors to interpolate transmit
beamformers, and is based on existing channel estimation
methods. Numerical results show that interpolation with the optimized
cluster size can perform significantly better than that with an
arbitrary cluster size. For the second proposed method, a channel
impulse response is quantized with a uniform scalar quantizer. 
With channel quantization, we also derive the
approximate sum achievable rate. We show that switching between the
two methods for different feedback-rate requirements can perform
better than the existing schemes.
\end{abstract}

\begin{IEEEkeywords}
Multiple-input single-output (MISO), OFDM, transmit beamforming,
feedback, RVQ, beamforming interpolation, optimal cluster size,
channel quantization.
\end{IEEEkeywords}

\section{Introduction}

Equipping a transmitter and/or a receiver with multiple antennas
creates a multiantenna wireless channel whose capacity depends on the
channel information available at the transmitter and/or receiver. In
multiantenna channels, transmit beamforming has been shown to increase
an achievable rate by directing transmit signal toward the strongest
channel mode~\cite{lo99}. With channel information, the receiver can
compute the optimal beamforming vector that maximizes achievable rate
and feeds the vector back to the transmitter. Due to a finite feedback
rate, the beamforming vector needs to be quantized. Several
quantization schemes and codebooks have been proposed and analyzed,
and the corresponding performance was shown to depend on the codebook
design and the number of available feedback bits~\cite[see references
  therein]{mimo, love06}. In this work, we consider transmit
beamforming for multiple-input single-output (MISO) orthogonal
frequency-division multiplexing (OFDM).

In MISO-OFDM, a wideband channel is converted into parallel narrowband
subchannels. For each subchannel or subcarrier, the optimal
beamforming vector is different and needs to be quantized at the
receiver and fed back to the transmitter. The total number of feedback
bits required increases with the number of subcarriers, which can be
large.  References~\cite{ClusterWu, choi, he11, huang11, ye, long12,
  pande, huang08, kim10, he12, chang12, wang12, zhou06, zhao12,
  ghirmai14} have proposed to reduce the amount of feedback
while maintaining performance. Due to high channel correlation in time
and frequency domains, feedback of transmit precoding matrices across
time and subcarriers can be compressed.  References~\cite{zhou06,
  zhao12} proposed to compress feedback with either recursive or
trellis-based encodings.

In~\cite{ClusterWu,long12,wang12,ghirmai14}, the optimal transmit
beamforming vectors of selected subcarriers, which are a few
subcarriers apart, are quantized while the remaining ones are
approximated to equal the quantized vector of the closest
subcarrier. The remaining transmit beamforming vectors are proposed
to be linearly interpolated in~\cite{choi,kim10} and spherically
interpolated in~\cite{he11,he12,huang08}. In~\cite{pande}, the
authors proposed to quantize the averaged optimal transmit
beamformer in each cluster termed mean clustering.  Geodesic-based
interpolation of transmit precoding matrices was also proposed
in~\cite{pande} and was extended to multiuser channels
in~\cite{chang12}. In~\cite{ye}, each subcarrier cluster uses the
same beamforming vector or precoding matrix, which is searched from
a subcodebook that contains entries close to the beamformer or
precoder in the adjacent cluster. Hence, there is some saving in
feedback bits.  Most of the works mentioned proposed to use either
the same or interpolated beamforming vectors for a group or cluster
of adjacent subcarriers since subcarriers are highly correlated in a
frequency-selective channel. However, none has analyzed the optimal
cluster size and the associated performance.

Given a limited feedback rate, we propose to quantize the optimal
beamforming vector at every few subcarriers with the random vector
quantization (RVQ) codebook proposed by~\cite{mimo} and to either use
the same quantized vector for the whole subcarrier cluster or
interpolate the remaining beamforming vectors in the cluster from the
quantized vectors. For the first proposed method termed constant
interpolation, we derive an approximate sum achievable rate over all
subcarriers. The analytical approximation can predict the performance
trend well and the optimal cluster size accurately. The optimal
cluster size depends mainly on the available feedback rate, and on how
frequency-selective the current channel is.

For linear interpolation, we propose a closed-form expression for
the phase-rotation parameter based on the correlation between the
transmit beamformers of subcarriers in the cluster. In earlier work
by~\cite{choi}, the parameter was exhaustively searched. Our
modified linear interpolation requires fewer minimum feedback bits
than that in~\cite{choi}. In~\cite{kim10}, the expression for the
phase-rotation parameter was also proposed and is based on a chordal
distance between two quantized beamforming vectors of the adjacent
clusters. However, our proposed phase rotation combined with the
optimized cluster size outperforms the phase rotation proposed
by~\cite{kim10}.

For higher-order interpolation, our method is based on earlier works
on comb-type pilot based channel estimation in
OFDM~\cite{liu92,hsieh98,he09}. Three or more quantized beamforming
vectors from adjacent clusters are used to interpolate all beamforming
vectors in one cluster. The number of phase-rotation parameters
increases with the order of the interpolation. The set of
phase-rotation parameters that maximizes sum achievable rate in a
cluster can be searched from the codebook proposed by~\cite{choi}.
Reference~\cite{he11} modified the second-order channel estimation
in~\cite{he09} to interpolate transmit beamforming vectors for
subcarriers in a multiple-input multiple-output (MIMO)-OFDM channel.
In~\cite{he12} the authors based their method from the work
by~\cite{liu92} to interpolate transmit beamformers for subcarriers,
but without phase rotations. The lack of phase-rotation parameters
degrades significantly the performance of the method in~\cite{he12}.
Our numerical example shows that the higher-order interpolation with
our optimized cluster size results in a good performance in a high
feedback-rate regime.

When the feedback rate is high, we propose to quantize the channel
impulse response with a uniform scalar quantizer and derive the
approximate sum rate for MISO channels. The scalar quantization used
in the proposed method is less complex than the vector quantization
used in~\cite{huang11}. The proposed scalar quantization of the
channel impulse response is shown to perform well with a high feedback
rate. Similar results were observed by~\cite{mimo} where the optimal
beamformer and not the channel response was scalar quantized. We note
that~\cite{diallo13} also proposed to scalar quantize a channel impulse
response, but the resulting sum rate was not analyzed.

Apart from what was presented earlier in~\cite{ecti12,icc13}, here we
show details of all proofs and update the derivation of the achievable
rate approximation in Proposition~\ref{t1}. We compare our proposed
feedback methods with several existing ones in the literature and show
that selecting the optimal cluster size that maximizes the sum rate
can significantly improve the sum rate. Higher-order interpolation
with quantized transmit beamforming is also proposed.

This paper is organized as follows. Section~\ref{sys_mod} describes
channel and feedback models as well as formulates the finite
feedback-rate problem. We propose beamforming interpolation methods
and analyze the optimal cluster size in Section~\ref{interp}. Direct
quantization of channel impulse response and its performance analysis
are shown in Section~\ref{quanch}. Numerical results and conclusions
are in Sections~\ref{num_re} and~\ref{conclude},
respectively. Finally, all proofs are in appendices.

\section{System Model}
\label{sys_mod}

We consider a point-to-point, discrete-time, MISO-OFDM channel with
$N$ subcarriers. A transmitter is equipped with $N_t$ antennas while a
receiver is equipped with a single antenna.  We assume that the
transmit antennas are placed sufficiently far apart that they are
independent.  For each transmit-receive antenna pair, a transmitted
signal propagates through a frequency-selective Rayleigh fading
channel with order $L$. Applying a discrete Fourier transform (DFT),
the frequency response for the $n$th subcarrier and the $n_t$th
transmit antenna is given by
\begin{equation}
   h_{n,n_t} = \sum_{l = 0}^{L-1} g_{l,n_t}
   \mathrm{e}^{\frac{-j2\pi ln}{N}}
\label{fft}
\end{equation}
where $g_{l,n_t}$ is a complex channel gain for the $l$th path between
the $n_t$th transmit and receive antenna pairs. Assuming a rich
scattering, $g_{l,n_t}$ for all $L$ paths and all $N_t$ transmit
antennas are independent complex Gaussian distributed with zero mean
and variance $E|g_{l,n_t}|^2$. In this work, we assume a uniform power
delay profile for which the power of each path is the same and the
total channel power for each transmit-receive antenna pair is
one. Hence, $E|g_{l,n_t}|^2 = \frac{1}{L}$.  Let $\bh_n$ denote an
$N_t \times 1$ channel vector of the $n$th subcarrier, whose entry is
$h_{n,n_t}$ shown in~\eqref{fft}. Thus,
\begin{equation}
  \bh_n = \left[ h_{n,1} \quad h_{n,2} \ \cdots \ h_{n,N_t}
    \right]^T .
\end{equation}

Assuming a transmit beamforming or a rank-one precoding, the
received signal on the $n$th subcarrier is given by
\begin{equation}
r_n = \bh_n^{\dag} \bv_n x_n + z_n, \quad 1 \le n \le N,
\label{received}
\end{equation}
where $\bv_n$ is an $N_t \times 1$ unit-norm beamforming vector, $x_n$
is a transmitted symbol with zero mean and unit variance, and $z_n$ is
an additive white Gaussian noise with zero mean and variance
$\sigma^2_z$. With perfect channel information at the transmitter, the
optimal transmit precoding that maximizes an achievable rate for MISO
channel is rank-one. This fact motivates us to use a rank-one
precoding or beamforming.  A resulting sum achievable rate over $N$
subcarriers is given by
\begin{equation}
C = \sum_{n = 1}^{N} E \left[\log(1 + \rho |\bh_n^{\dag}
\bv_n|^2)\right] \label{c}
\end{equation}
where the expectation is over the distribution of $\bh_n$. We assume a
uniform power allocation for all subcarriers and hence, the
background signal-to-noise ratio (SNR) for each subcarrier $\rho =
1/\sigma^2_z$.

From~\eqref{c}, we note that the sum achievable rate is a function
of transmit beamforming vectors $\{\bv_1, \bv_2, \ldots, \bv_N\}$. A
receiver with perfect channel information can optimize the sum
achievable rate over the transmit beamforming vectors and send the
selected beamforming vectors to the transmitter via a feedback
channel. Since the feedback channel between the receiver and the
transmitter has a finite rate, quantizing the transmit beamforming
vectors is required. In this study we apply a random vector
quantization (RVQ) codebook whose entries are independent,
isotropically distributed vectors to quantize a transmit beamforming
vector. RVQ is simple, however has been shown to perform close to
the optimum codebook~\cite{mimo,yeung_miso}.

We assume $B$ total feedback bits per update. For an
equal-bit-per-subcarrier allocation, each beamforming vector is
quantized with $B/N$ bits. Let us denote the RVQ codebook by $\sV =
\{\bw_1, \bw_2, \ldots, \bw_{2^{B/N}} \}$ with $2^{B/N}$ entries. The
receiver selects for the $n$th subcarrier the entry in the codebook
that maximizes an instantaneous achievable rate as follows:
\begin{align}
  \bvh_n & = \arg \max_{\bw \in \sV} \ \log (1 + \rho |\bh_n^{\dag}
                  \bw|^2 ) \\
             & = \arg \max_{\bw \in \sV} \ |\bh_n^{\dag}\bw|^2
\label{quant}
\end{align}
and the associated achievable rate for the $n$th subcarrier is given
by
\begin{align}
  C_n &= E \left[ \log ( 1 + \rho |\bh_n^{\dag}
  \bvh_n|^2 ) \right]\\
  &= E \left[ \log ( 1 + \rho \|\bh_n\|^2|\bhb_n^{\dag}
  \bvh_n|^2 ) \right]. \label{cap_rvq}
\end{align}
where $\bhb_n= \bh_n/\|\bh_n\|$ is a unit-norm channel vector that
points in the same direction as $\bh_n$. Evaluating~\eqref{cap_rvq}
was shown by~\cite{yeung_miso}. We note from~\eqref{cap_rvq} that the
achievable rate depends on the number of feedback bits per subcarrier,
which could be small due to a large number of subcarriers in a
practical OFDM system. Hence, this may result in a large quantization
error, which leads to a substantial performance loss.

\section{Interpolating Transmit Beamforming Vectors}
\label{interp}

Feeding back transmit beamforming vectors of all subcarriers requires
quantizing $NN_t$ complex coefficients and thus, a large number of
feedback bits. We note that adjacent subcarriers in OFDM are highly
correlated since the number of channel taps is much lower than that of
subcarriers ($L \ll N$). The optimal transmit beamformers, which
depend on channel matrices, are also highly correlated. In this
section, we propose beamforming interpolation of different orders to
reduce the number of feedback bits while maintaining the performance.

First we evaluate a squared correlation between
normalized channel vectors of subcarrier $n$ and $n+q$ defined by
\begin{equation}
E\left[ |\bhb_n^{\dag}\bhb_{n+q}|^2 \right] =
E\left[\frac{|\bh_n^{\dag}\bh_{n+q}|^2}{\|\bh_n\|^2\|\bh_{n+q}\|^2}\right].
    \label{afH}
\end{equation}
Evaluating~\eqref{afH} is not tractable for a finite-size
system. Hence, we approximate the average squared correlation as follows.
\begin{lemma}
\label{lemma_cor}
A squared correlation between the $n$th and $n+q$th normalized channel vectors
is approximated as follows:
\begin{align}
  E\left[ |\bhb_n^{\dag}\bhb_{n+q}|^2 \right] &\approx \frac{L^2 + N_t
    \varphi^2(q)}{L^2N_t + \varphi^2(q)} \label{baL} \\ &\triangleq
  \psi(q,N_t) \label{afN}
\end{align}
where
\begin{equation}
\varphi(x) = \frac{\sin(\frac{\pi
  xL}{N})}{\sin(\frac{\pi x}{N})} .
\end{equation}
\end{lemma}
The proof of Lemma~\ref{lemma_cor} is shown in
Appendix~\ref{append_lemma_cor}.

As subsequent numerical example in Section~\ref{num_re} will show that
approximation in Lemma~\ref{lemma_cor} closely predicts the result of
a finite-size system. The correlation in~\eqref{baL} depends on $L$,
$N$, $N_t$, and most importantly, $q$, which indicates how far apart
the two channel vectors are. When $q \to 0$, $\varphi(q) \to L$ and
the squared correlation becomes $\|\bhb_{n}\|^4 \to 1$. We note that
the number of channel taps $L$ and channel impulse response can be
accurately estimated as shown in~\cite{borah99}.

\subsection{Constant Interpolation}
\label{Sub1}

In the first method, we group adjacent contiguous subcarriers into a
cluster and apply the same quantized beamforming vector for all
subcarriers in the cluster. We denote the number of contiguous
subcarriers in one cluster by $M$.  Thus, the number of clusters is
given by $K \triangleq \lfloor N/M \rfloor$ with a possible few
remaining subcarriers. The number of feedback bits allocated for each
cluster is equal to $B/K$. All $B/K$ bits are used to quantize the
beamforming vector of the centered subcarrier for odd $M$ and one
subcarrier off the center for even $M$. Therefore, the beamforming
vector used for the $k$th cluster is given by
\begin{equation}
\bvh_{kM+m}  =  \left\{ \begin{array}{lr} \arg \max_{\bw \in \sV} \
|\bhb_{kM + \frac{M+1}{2}}^{\dag} \bw |^2 & \text{for odd} \ M\\
\arg \max_{\bw \in \sV} \ |\bhb_{kM + \frac{M}{2}}^{\dag} \bw |^2 &
\text{for even} \ M \end{array} \right. \label{NoInter}
\end{equation}
where $1 \le m \le M$ and $0 \le k \le K - 1$. If $N/M$ is not
an integer, then there exist some remaining subcarriers, which do not
belong in any cluster. We propose to set the transmit beamforming for
these subcarriers to be that of the last cluster as follows:
\begin{equation}
\bvh_{K M + q} = \bvh_{K M} \quad \text{for} \ 1 \le q \le N - K M .
\end{equation}

With constant interpolation, an achievable rate for each subcarrier
can be approximated by Proposition~\ref{t1} .

\begin{prop}
\label{t1} For $1 \le n+q \le N$, the approximate ergodic
achievable rate of the $(n+q)$th subcarrier is given by
\begin{equation}
 C_{n+q} \approx \mathsf{C}_{n+q} =  \log(1 +\rho N_t \gamma(n+q,
 B/K))
\end{equation}
where
\begin{multline}
\gamma(n+q, B/K) \triangleq \psi(q, N_t) \cdot(1- 2^{B/K}
\beta(2^{B/K}, \frac{N_t}{N_t-1})) \\ + (1 -\psi(q,N_t)) \cdot
\frac{(2^{B/K}\beta(2^{B/K},\frac{N_t}{N_t-1}))}{N_t-1}
\label{Gram}
\end{multline}
and the beta function $\beta(x, y) = \int_0^1 t^{x-1}(1-t)^{y-1} \,
\diff t$.
\end{prop}

The proof of Proposition~\ref{t1} is shown in Appendix~\ref{append_prop}.

With Proposition~\ref{t1}, we obtain the approximate sum achievable
rate for a single cluster with odd $M$ as follows:
\begin{multline}
\mathsf{C}_{\text{cluster}} = \sum_{q = -
  \frac{M-1}{2}}^{\frac{M-1}{2}} \mathsf{C}_{n+q} = \log (1 + \rho N_t
\gamma(0,B/K)) \\
+ 2 \sum_{q=1}^{\frac{M-1}{2}} \log (1 + \rho N_t
\gamma(q,B/K)) .
\label{odd}
\end{multline}
With even $M$, the sum achievable rate for a single cluster is
approximated by
\begin{multline}
\mathsf{C}_{\text{cluster}} = \log (1 + \rho N_t \gamma(0,B/K))
\\ + 2 \sum_{q=1}^{\frac{M}{2} - 1} \log (1 + \rho N_t \gamma(q,B/K))
+  \log (1 + \rho N_t \gamma(\frac{M}{2}, B/K)) .
\label{even}
\end{multline}

We note that the performance of the constant interpolation has a
trade-off between total feedback bits and cluster size and hence,
there exists optimal cluster size for a given feedback budget. Given
$B$ feedback bits and other system parameters, we would like to
determine the number of subcarriers $M^*$, which maximizes the
approximate sum achievable rate of {\em all} $N$ subcarriers given
by
\begin{multline}
   M^* = \arg \max_{\substack {1 \le M \le N \\M \in \mathbb{Z}}}
   \Bigg\{ K \mathsf{C}_{\text{cluster}} \\+ \sum_{r = 1}^{N - K M} \log
   (1 + \rho N_t \gamma(r + \frac{M}{2}, B/K)) \Bigg\}
\label{N1}
\end{multline}
where the first term accounts for the approximate achievable rate of
the $K$ clusters and the second term accounts for the approximate
achievable rate of a few remaining subcarriers. Solving~\eqref{N1} can
be accomplished by either integer programming for which there exist
many available tools or by exhaustive search. Although the
optimization in~\eqref{N1} is based on the approximation of the actual
achievable rate, subsequent numerical examples in Section~\ref{num_re}
show that the solution to~\eqref{N1} accurately predicts the optimal
cluster size. We note that there is no other comparable analysis on
the optimal cluster size in the literature.

Besides the sum achievable rate, another important performance metric
is the average received power across subcarriers defined as follows:
\begin{align}
\eta_{\text{AVE}} &\triangleq \frac{1}{N}\sum_{n=1}^N \rho E
\left[|\bh_n^{\dag}\bvh_n|^2\right] \\
& = \frac{\rho N_t}{N}\sum_{n=1}^{N} E\left[|\bhb_{n}^\dag\bvh_n|^2\right].
\end{align}
where it is shown in the proof of Proposition~\ref{t1} that
\begin{equation}
  E\left[|\bhb_{n}^\dag\bvh_n|^2\right] \approx \gamma(n, B/K).
\end{equation}

Therefore, the average received power can be approximated as follows.
\begin{corollary}
For odd $M$,
\begin{multline}
\eta_{\text{AVE}} \approx \frac{\rho N_t}{N} \Bigg\{K \gamma(0,B/K)
+ 2K \sum_{q=1}^{\frac{M-1}{2}} \gamma(q,B/K) \\+ \sum_{r = 1}^{N
- K M} \gamma(r + \frac{M}{2}, B/K) \Bigg\}, \label{oddp}
\end{multline}
and for even $M$
\begin{multline}
\eta_{\text{AVE}} \approx \frac{\rho N_t}{N} \Bigg\{ K \gamma(0,B/K) +
2K \sum_{q=1}^{\frac{M}{2} - 1} \rho \gamma(q,B/K) \\+ K
\gamma(\frac{M}{2}, B/K) + \sum_{r = 1}^{N - K M}
\gamma(r + \frac{M}{2}, B/K) \Bigg\}. \label{evenp}
\end{multline}
\label{c1}
\end{corollary}
These analytical expressions give a more accurate approximation than
those in Proposition~\ref{t1} since there is no Jensen's inequality
involved as demonstrated by numerical results in
Section~\ref{num_re}.

\subsection{Linear Interpolation}
\label{Sub2}

To increase the performance, we propose to modify a linear
interpolation proposed by~\cite{choi}. Similar to the constant
interpolation, all subcarriers are grouped into $K$ clusters. Each
cluster consists of $M$ contiguous subcarriers and a possible last
cluster with a few remaining subcarriers. For each cluster, the
optimal beamforming vector of the first subcarrier is selected from an
RVQ codebook with either $B/K$ bits or $B/(K+1)$ bits, depending on
the total number of clusters.

All other beamforming vectors in a cluster are linear combinations of the
quantized beamforming vector of the first subcarrier in the cluster
and that in the next cluster as follows~\cite{choi}:
\begin{equation}
  \bvh_{kM + m}(\theta_m) \triangleq \frac{(1-c_m) \bvh_{kM}+c_m \me^{j\theta_m}
    \bvh_{(k+1)M}}{\|(1-c_m)\bvh_{kM} + c_m
    \me^{j\theta_m}\bvh_{(k+1)M}\|}
\label{intv}
\end{equation}
for $1 \le m \le M-1$ and $0 \le k \le K-1$, where
\begin{equation}
c_m = \frac{m}{M}
\label{eq_cm}
\end{equation}
is a linear weight and $\theta_m$ is a phase-rotation parameter. We
note that for the last cluster, we choose to interpolate with $\bvh_1$
instead of $\bvh_N$ to save some feedback bits. Due to DFT, $\bhb_1$
is similar to $\bhb_N$ and hence, $\bvh_1$ is also similar to $\bvh_N$.

In~\cite{choi}, $\theta_m$ is chosen to maximize the sum achievable
rate in~\eqref{c} by performing exhaustive search over the received
power in each cluster as follows. For the $k$th cluster,
\begin{equation}
\theta_m = \arg \max_{\theta \in \Theta} \sum_{i = 1}^{M}
|\bh_{kM+i}^{\dag}\bvh_{kM+i}(\theta)|^2 \label{Qtp}
\end{equation}
where the phase-rotation codebook
\begin{equation}
\Theta = \left\{ 0, 2\pi\frac{1}{P}, 2\pi\frac{2}{P}\ldots, 2\pi
\frac{P-1}{P} \right\}
\end{equation}
and $P$ is the number of quantization levels.

To avoid search complexity and reduce feedback, here we propose to
determine the phase rotation based on a correlation between the
optimal beamformers of neighboring subcarriers.  We note that the
optimal transmit beamforming vector for the $n$th subcarrier is
matched to the normalized channel vector $\bv_n^{\opt} =
\bhb_n \label{di}$. Evaluating a correlation between the optimal
beamformer and the interpolated beamformer that are $m$ subcarriers
apart, $E|(\bv_{kM}^{\opt})^{\dag}\bv_{kM+m}|^2$, follows similar
steps to the proof of Lemma~\ref{lemma_cor}. This correlation is most
likely close to the correlation between the optimal beamformers, which
is approximated to be $\psi(m, N_t)$ in~\eqref{afN}. Based on this
assumption, we set $E|(\bv_{kM}^{\opt})^{\dag}\bv_{kM+m}|^2$ to
$\psi(m, N_t)$ and solve for the phase-rotation parameter given by the
following proposition.
\begin{prop}
\label{c2}
 The phase rotation for the $m$th subcarrier in the cluster is given
 by
\begin{equation}
  \theta_m = \arccos \frac{U(m)}{V(m)}
\label{ttm}
\end{equation}
where
\begin{multline}
 U(m) = (1-c_m)^2(\psi(m,N_t) - N_t + 1)
   \\+ c_m^2(N_t\psi(m,N_t) - \frac{N_t}{L^2}\varphi^2(M) + 1)
\end{multline}
and
\begin{multline}
 V(m) = \frac{2}{L} (1-c_m) c_m (N_t - N_t\psi(m,N_t) + 1) \\
    \cdot \varphi(M)
  \cos\left(\frac{\pi M (L-1)}{N}\right).
   \label{Vm}
\end{multline}
\end{prop}
The proof is shown in Appendix~\ref{pp2}.

Finding the optimal cluster size for the linear interpolation is not
tractable since the achievable rate expression is not known. From
numerical results, the optimal cluster of the constant interpolation
mostly aligns with that of the linear interpolation and that of
higher-order interpolations as well. We note that computing $\theta_m$
in Proposition~\ref{c2} can be performed at the transmitter with the
number of channel taps $L$ and cluster size $M$. For a relatively
static environment, $L$ and hence $M$ may not change
often~\cite{nguyen07}. Thus, feedback for these parameters do not
occur often and consists of minimal number of bits. For~\cite{choi},
the phase-rotation needs to be fed back for every cluster. The number
of additional feedback bits in~\cite{choi} increases linearly with the
number of clusters and can be significantly larger than that in our
method.

\subsection{Higher-Order Interpolation}

For a better interpolation, more than two quantized transmit
beamforming vectors should be used to interpolate the beamforming
vectors in the cluster. For instance, the second-order interpolated
transmit beamformer in the $k$th cluster is as follows:
\begin{multline}
   \bvh_{kM + m} \\= \frac{\alpha_{-1} \me^{j\theta_{m;-1}}
     \bvh_{(k-1)M} + \alpha_0 \bvh_{kM} + \alpha_1 \me^{j\theta_{m;1}}
     \bvh_{(k+1)M}}{\| \alpha_{-1} \me^{j\theta_{m;-1}} \bvh_{(k-1)M}
     + \alpha_0 \bvh_{kM} + \alpha_1 \me^{j\theta_{m;1}} \bvh_{(k+1)M}
     \|} . \label{2oder}
\end{multline}
We note that there are 3 quantized beamformers $\bvh_{(k-1)M},
\bvh_{kM}, \bvh_{(k+1)M}$, which are used for interpolation. This
interpolation was modified from the channel interpolation methods
proposed by~\cite{hsieh98,liu92}. The set of constants is given
by~\cite{liu92}
\begin{gather}
  \alpha_{-1} = \frac{1}{2}c_m (c_m-1)\\
   \alpha_0 = -(c_m-1)(c_m+1)\\
   \alpha_1 = \frac{1}{2}c_m ( c_m+1) .
\label{C2nd}
\end{gather}
Phase-rotation parameters $\theta_{m;-1}$ and $\theta_{m;1}$ are
introduced in this study to increase the performance of the
higher-order interpolation. Similar to that in the linear
interpolation, the set of the two phase rotations is found by
maximizing the sum received power in the $k$th cluster over the
codebook $\Theta$ as follows:
\begin{equation}
\max_{\{\theta_{m;-1}, \theta_{m;1} \} \in \Theta^2} \sum_{i = 1}^{M} |\bh_{kM+i}^{\dag}\bvh_{kM+i}|^2 .
\label{QtpO}
\end{equation}

For order $R$ where $R$ is even and $R > 2$, the interpolated
beamformer in the $k$th cluster is given by
\begin{equation}
  \bvh_{kM+m} = \frac{\bf{y}}{\|\bf{y}\|}
\end{equation}
where 
\begin{multline}
  {\bf{y}} = \sum_{s = -\frac{R}{2}}^{-1}
    \alpha_s\me^{j\theta_{m;s}}\bvh_{(k+s)M} \\+ \alpha_0\bvh_{kM} + \sum_{t = 1}^{\frac{R}{2}} \alpha_t \me^{j\theta_{m;t}}\bvh_{(k+t)M} 
\end{multline}
and $\{\alpha_r\}_{r=-\frac{R}{2}}^{\frac{R}{2}}$ is a set of
interpolation constants while the set of phase rotations
$$\{\theta_{m;-\frac{R}{2}}, \ldots, \theta_{m;-1}, \theta_{m;1},
\ldots, \theta_{m;\frac{R}{2}}\}$$ can be found by exhaustive search
over the phase codebook $\Theta$.

We expect the higher-order interpolation method to perform better than
the previous methods, but the performance gain is obtained at the
expense of additional complexity and feedback. Search complexity to
locate the optimal set of phase-rotation parameters increases with the
number of phase rotations or the order of the interpolation. Also, the
additional number of feedback bits to quantize these phase rotations
increases with the number of clusters and the order of the
interpolation. These bits are in addition to the number of bits used
to quantize transmit beamformers.

\section{Quantizing Channel Impulse Response}
\label{quanch}

When the available feedback rate is sufficiently high (larger than 2
bits per complex entry), quantizing the channel impulse response
directly can perform well~\cite{commag04}. Here we propose to quantize
all channel taps of all transmit-receive antenna pairs with a scalar
uniform quantizer. A uniform quantizer is simple and performs close to
the optimal quantizer when the number of quantization bits is
high. Real and imaginary parts of all channel taps are quantized
independently with the same number of bits, which is
$\frac{B}{2N_tL}$.  Thus, the quantized $l$th channel tap for the
$n_t$th transmit-receive antenna pair is given by
\begin{align}
  \hat{g}_{l, n_t} & = \hat{g}_{l, n_t, r} +
  j\hat{g}_{l, n_t, i}\\
  &= Q (g_{l, n_t, r}) + jQ(g_{l, n_t, i})
\end{align}
where $g_{l, n_t, r}$ and $g_{l, n_t, i}$ are real and imaginary parts
of $g_{l, n_t}$, respectively, $Q(\cdot)$ is the uniform scalar
quantizer with $2^{\frac{B}{2N_tL}}$ steps, while variables with hats
denote outputs of the quantizer.  Here we select a step size of the
quantizer by the existing rule of thumb for Gaussian input
(cf.~\cite[p. 125]{jayant84})
\begin{equation}
  \Delta = \frac{4E[(g_{l, n_t, r})^2]}{2^{\frac{B}{2N_tL}}}
  = \frac{1}{\sqrt{L}} 2^{\frac{3}{2}-\frac{B}{2N_tL}} ,
\end{equation}
which changes with the variance of the channel tap and the number of
quantization bits.  Then, the transmitter computes a DFT of the
quantized channel impulse response to obtain an approximate frequency
response as follows:
\begin{equation}
 \hat{h}_{n, n_t} =
 \sum_{l=0}^{L-1} \hat{g}_{l, n_t} \me^{-\frac{j2\pi l n}{N}} ,
\end{equation}
which is the $n_t$th entry of the quantized $N_t \times 1$ channel
vector for the $n$th subcarrier denoted by $\hat{\bh}_n = \left[
  \hat{h}_{n, 1} \quad \hat{h}_{n, 2} \ \cdots \ \hat{h}_{n, N_t} \right]^T$.

Based on $\hat{\bh}_n$, the transmitter transmits signal in the
direction of the quantized channel vector, namely,
$\hat{\bh}_n/\|\hat{\bh}_n\|$ and the corresponding sum rate over all
subcarriers is given by
\begin{align}
C &= \sum_{n = 1}^{N} E\left[\log ( 1 + \rho \frac{|\bh_n^\dag \hat{\bh}_n|^2}
    {\|\hat{\bh}_n\|^2})\right] \label{sn1}\\
  &= N E\left[\log ( 1 + \rho \frac{|\bh_n^\dag \hat{\bh}_n|^2}
    {\|\hat{\bh}_n\|^2})\right] \label{sn2} \\
  &\le N \log ( 1 + \rho E \left[ \frac{|\bh_n^\dag \hat{\bh}_n|^2}
    {\|\hat{\bh}_n\|^2} \right]) \label{sn3}\\
  &\approx N \log ( 1 + \rho \frac{E [|\bh_n^\dag \hat{\bh}_n|^2]}{E
    [\|\hat{\bh}_n\|^2]}) \label{sn4}
\end{align}
where in~\eqref{sn2}, we use the fact that the distribution of
subcarriers is identical and in~\eqref{sn3} and~\eqref{sn4}, we apply
Jensen's inequality and approximate an expectation of the quotient by
a quotient of the two expectations. The approximation becomes more
accurate as the number of transmit antennas
increases~\cite{zhang14}. Consequently, we obtain the approximate sum
achievable rate.

Since real and imaginary parts of each channel tap are independent
and Gaussian distributed with zero mean and variance $\frac{1}{2L}$, we
can easily show that
\begin{equation}
  E [\|\hat{\bh}_n\|^2] = N_t(1 - 2L E[(\hat{g}_{r} - g_r)^2])
\label{nt1}
\end{equation}
and
\begin{equation}
\begin{split}
  E [|\bh_n\hat{\bh}_n^\dag|^2] = N_t(&1 + \frac{1}{L} - (2L -
  1)E[(\hat{g}_{r} - g_r)^2] \\
  &+ 2L E[ \hat{g}_r^2 g_r^2] +
  4L(N_tL-1)E^2[\hat{g}_r g_r])
\end{split}
\label{2le}
\end{equation}
where we have dropped indices $n_t$ and $l$ from $g_{l, n_t, r}$ for
clarity.  The mean squared error is given by
\begin{equation}
  E[(\hat{g}_{r} - g_r)^2] = \int (Q(x) - x)^2 f_{g_r} (x) \,
  \mathrm{d} x \label{Ehh}
\end{equation}
and the correlation and its second moment are given by
\begin{gather}
  E [\hat{g}_r g_r] = \int x Q(x)f_{g_r} (x) \, \mathrm{d} x\\
  E[ \hat{g}_r^2 g_r^2] = \int x^2 Q^2(x)f_{g_r} (x) \, \mathrm{d} x .
\label{Ehr}
\end{gather}
where $f_{g_r} (\cdot)$ denotes the probability density function (pdf)
of $g_{l, n_t, r}$.

Each term in~\eqref{Ehh}-\eqref{Ehr} can be computed numerically.
However, to obtain some insight on how the sum achievable rate
depends on the feedback rate and other channel parameters, we
approximate each term in a high feedback-rate regime. It was shown
that for large $B$~\cite{hui01},
\begin{equation}
  E[(\hat{g}_{r} - g_r)^2] \approx \frac{\Delta^2}{12} =
  \frac{2}{3L}2^{-\frac{B}{N_tL}}.
\label{Btf}
\end{equation}
Applying the property of the optimum quantizer~\cite{bucklew79}, we
obtain
\begin{equation}
   E [\hat{g}_r g_r] \approx \frac{1}{2L} - E[(\hat{g}_{r} - g_r)^2] .
\end{equation}
As $B \to \infty$, $\hat{g}_r \to g_r$. Hence,
\begin{equation}
  \lim_{B \to \infty} E[ \hat{g}_r^2 g_r^2] = \frac{3}{4L^2} .
\label{Btfh}
\end{equation}

Substituting \eqref{Btf} -- \eqref{Btfh} into \eqref{nt1} and
\eqref{2le}, we obtain the approximate upper bound for a sum
achievable rate for the MISO channel with large $B$ as follows
\begin{equation}
  C \approx N \log (1 + \rho(1 - \frac{1}{L} + (N_tL-1)\Omega_B +
  \frac{3}{4 L^2 \Omega_B}))
\label{clsim}
\end{equation}
where $\Omega_B = \frac{1}{L} - \frac{4}{3L}2^{-\frac{B}{N_tL}}$. As
the number of feedback bits per transmit antenna and channel tap,
$\frac{B}{N_t L}$, increases, the quantization error~\eqref{Btf}
decreases and the sum rate~\eqref{clsim} increases. We note that for
as $\frac{B}{N_t L} \to \infty$ and $N_t$ increases, the sum rate
in~\eqref{clsim} increases as $N \log(\rho N_t)$, which is the sum
rate with perfect CSI. With a large feedback rate, quantizing channel
impulse response can achieve a larger sum rate than beamforming
interpolation as will be shown in subsequent numerical examples. To
determine at what $B$ to switch from channel quantization to, for
example, the constant beamforming interpolation, we compare the sum
rate obtained from Proposition~\ref{t1} with that
from~\eqref{clsim}.

\section{Numerical Results}
\label{num_re}

\begin{figure}[ht]
\centering
\includegraphics[width=3.5in]{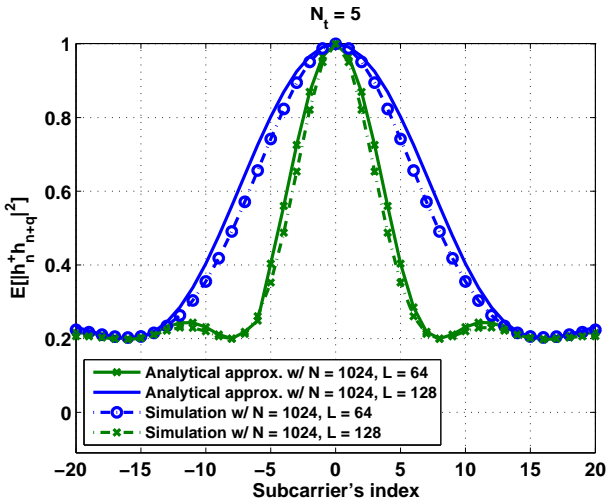}
\caption{Correlation between subcarriers $E|\bhb_n^\dag \bhb_{n+q}|^2$
  from both simulation and analytical results with $N_t = 5$, $N =
  1024$, and $L = 64 \text{ and } 128$.}
\label{fig1}
\end{figure}

To illustrate the performance of the proposed interpolations, Monte
Carlo simulation is performed with 3000 channel
realizations. Fig.~\ref{fig1} shows a correlation between subcarriers
$E|\bhb_n^\dag\bhb_{n+q}|^2$ from simulation results and the
analytical approximation in Lemma~\ref{lemma_cor} with $N_t = 5$, $N
=1024$, $L$ = 64 and 128, respectively. From this figure, we see that
the correlation between subcarriers decreases as expected when the
subcarriers are further apart and note that the analytical
approximation derived in Lemma~\ref{lemma_cor} predicts the simulation
results quite accurately.

\begin{figure}[ht]
\centering
\includegraphics[width=3.5in]{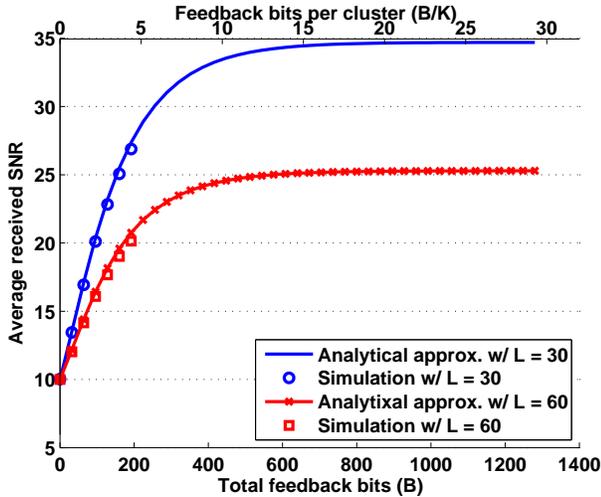}
\caption{Average received power per subcarrier $\eta_{\text{AVE}}$
  with the number of total feedback bits $B$ for $N = 1024$, $N_t =
  4$, $M = 32$, and SNR at $10$ dB.}
 \label{f3add}
\end{figure}

Fig.~\ref{f3add} shows the average received power per subcarrier
$\eta_{\text{AVE}}$ with constant interpolation for different
numbers of feedback bits $B$ and channel taps $L$. We set the number
of transmit antennas $N_t = 4$, cluster size $M = 32$, and SNR at 10
dB. We also place another x-axis on the top of the figure showing
the number of bits per cluster $B/K$. In the figure, the solid lines
show the analytical approximation given in Corollary~\ref{c1} while
the square and circular markers show the simulation results. Due to
search complexity of RVQ, there are no simulation results for a
large-feedback regime.

We note from the figure that the average received power increases with
$B$ as expected and decreases with $L$. As the channel becomes more
frequency selective, the cluster size should be reduced to maintain
the performance. We observe that in this example, the analytical
results are very close to those from the simulation. Unlike the
achievable rate analysis, Jensen's inequality is not used in deriving
$\eta_{\text{AVE}}$.  From the figure, we see that about half a
feedback bit per subcarrier gives us close to the infinite-feedback
performance.

\begin{figure}[ht]
\centering
\includegraphics[width=3.5in]{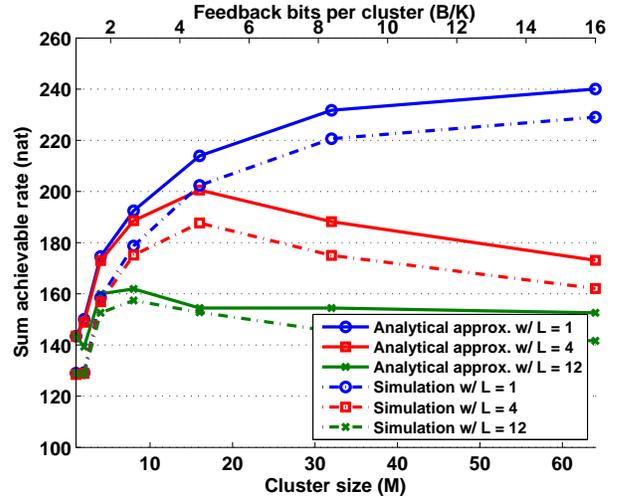}
\caption{Sum achievable rate with different cluster size $M$ and
  different number of channel taps $L$ for $N = 64$, $N_t = 4$, $B =
  16$, and SNR at 10 dB.}
 \label{fig5}
\end{figure}

Fig.~\ref{fig5} shows the sum achievable rate with constant
interpolation with cluster size $M$ from both the analytical
approximation from Proposition~\ref{t1} and the simulation results
when the number of total feedback bits is severely limited at 16
bits. Different plots correspond to different $L$ values. For small
$M$, more beamforming vectors are quantized and fed back from the
receiver, but with a smaller number of feedback bits per cluster.
For large $M$, the opposite is true. Thus, there exists an optimal
$M$ that maximizes the achievable rate. We can observe from this
figure, selecting optimal $M=16$ performs 35\% better than that for
feeding back every subcarrier ($M = 1$) for $L=4$. Comparing the
analytical approximation and the simulation results, we observe that
the gap is quite substantial (due to Jensen's inequality); however,
the analytical result still can accurately predict the optimal $M$.
For a flat fading channel ($L=1$), all subcarrier gains are the same
and thus, the optimal $M^*=1$. For frequency selective fading ($L
> 1$), subcarriers are less correlated and the optimal $M^*$
decrease with $L$. We remark that the system size for this figure is
set to be smaller than that in the previous figure. This is due
again to computational complexity of RVQ when $B$ is large.

\begin{figure}[ht]
\centering
\includegraphics[width=3.5in]{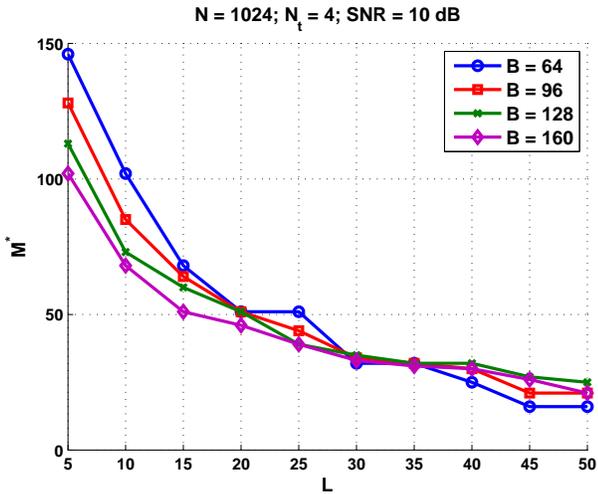}
\caption{The optimal $M^*$ shown with $L$ and $B$ for $N = 1024$, $N_t =
  4$, and SNR at 10 dB.}
 \label{1024_fig4}
\end{figure}

Fig.~\ref{1024_fig4} shows the optimal number of subcarriers per
cluster $M^*$ obtained from the analytical bound approximation with
different numbers of channel taps and total feedback bits. In this
figure, we observe that $M^*$ decreases when $L$ increases. In other
words, when the channel becomes more frequency selective, cluster size
should be reduced. Furthermore, with more available feedback bits,
cluster size should also be reduced. The explanation is as
follows. As shown in Fig.~\ref{f3add}, an increase in the number of
quantization or feedback bits beyond a certain point will give
diminishing rate return. Therefore, to extend a rate increase, cluster
size should be reduced for a better interpolation of transmit
beamforming.

\begin{figure}[ht]
\centering
\includegraphics[width=3.5in]{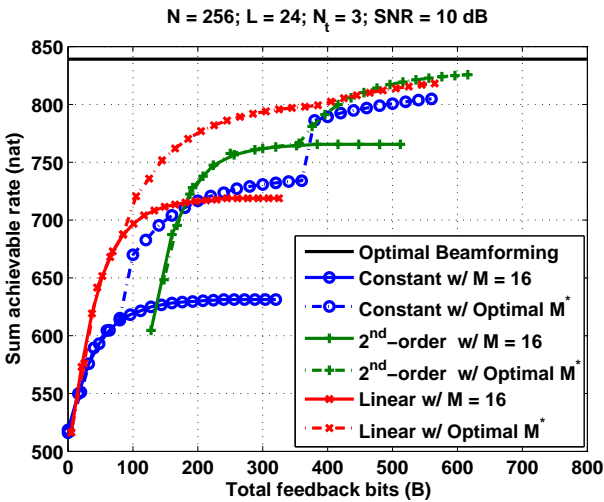}
\caption{Comparison of the sum achievable rate with different
  interpolation methods and with either optimized or fixed cluster
  sizes for $N = 256$, $L = 24$, $N_t = 3$, and SNR at $10$ dB.}
 \label{f1add}
\end{figure}

In Fig.~\ref{f1add}, we compare the sum achievable rate of all
interpolation methods proposed in the study with either the
optimized cluster size obtained from~\eqref{N1} or fixed cluster
size $M = 16$. For this figure, we set $N = 256$, $L = 24$, $N_t =
3$, and SNR at 10 dB.  We observe that with a low feedback,
the performance of interpolation with optimal cluster size and that
with $M = 16$ do not differ much. However,the performance gap
between interpolation with or without the optimized cluster size
widens significantly as available feedback becomes larger. The gain
on the performance gain could be as high as 15\% for the constant
interpolation. The solid line on the top of the figure shows the
infinite feedback performance. We see that with only two bits per
subcarriers, our proposed methods achieve near optimal performance.
We note that the second-order interpolation performs worse than the
constant or the first-order interpolation in low or moderate $B$
regimes. This is due to the extra feedback bits required to feed
back the two phase-rotation parameters $\theta_{m;
  -1}$ and $\theta_{m;1}$ by the second-order method. The additional
feedback bits can be significant. For a fixed $M = 16$ (hence, $K =
16$), 8 bits per cluster or total 128 bits are used to feed back the
two phase-rotation parameters.

\begin{figure}[ht]
\centering
\includegraphics[width=3.5in]{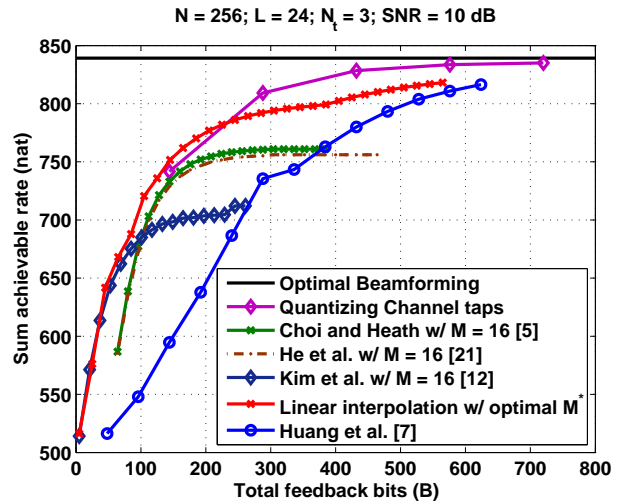}
\caption{Sum achievable rate of a $3 \times 1$ OFDM channel with
various
  feedback schemes plotted with the total number of feedback bits $B$
  and $N = 256$, $L = 24$, and SNR at 10 dB.}
 \label{f2add}
\end{figure}

In Fig.~\ref{f2add}, we compare our linear interpolation method with
the optimized subcarrier cluster size from Section~\ref{interp} and
direct channel quantization from Section~\ref{quanch} with existing
methods~\cite{choi,he09,kim10,huang11}. References~\cite{choi,he09,kim10}
propose a beamforming interpolation in frequency domain while
\cite{huang11} proposes to vector-quantize channel impulse
response. In both~\cite{choi,he09}, a single phase-rotation parameter
is used for the whole cluster. Hence, the number of phase rotations to
be quantized and fed back in~\cite{choi,he09} equals the number of
clusters. The main difference between~\cite{choi} and~\cite{he09} is
linear weight $c_m$\footnote{The expression of $c_m$ in~\eqref{eq_cm}
  was also used by~\cite{choi}.}.  In our method, the phase rotation
$\theta_m$ differs for different subcarriers in the same cluster and
can be determined at the transmitter with just the number of channel
taps $L$ and cluster size $M$ fed back to the transmitter. Thus, the
additional number of feedback bits in our method is minimal while
those in~\cite{choi,he09} increase linearly with the number of
clusters $K$. For this figure, methods proposed by~\cite{choi,he09}
require 64 additional bits. In~\cite{kim10}, phase rotation is based
on the chordal distance between the two adjacent quantized beamformers
and is the same for all subcarriers in the cluster.
References~\cite{choi,he09,kim10} do not optimize cluster size and in
this figure, it is fixed at 16. For~\cite{huang11}, magnitudes and
phases of all channel taps are vector-quantized. The method proposed
in~\cite{huang11} performs worse than our channel quantization for
small $B$. However, we expect the performance of the two methods to be
comparable when $B$ is large.

From Fig.~\ref{f2add}, we remark that the combination of our methods
(linear interpolation in a low feedback-rate regime and direct channel
quantization in a high feedback-rate regime) dominates all mentioned
works in all feedback range. Also from this figure, we can conclude
that with roughly one feedback bit per subcarrier, the direct
channel-tap quantization is preferred, and with fewer than one bit per
subcarrier, interpolation from quantized transmit beamformers is
preferred.

\begin{figure}[ht]
\centering
\includegraphics[width=3.5in]{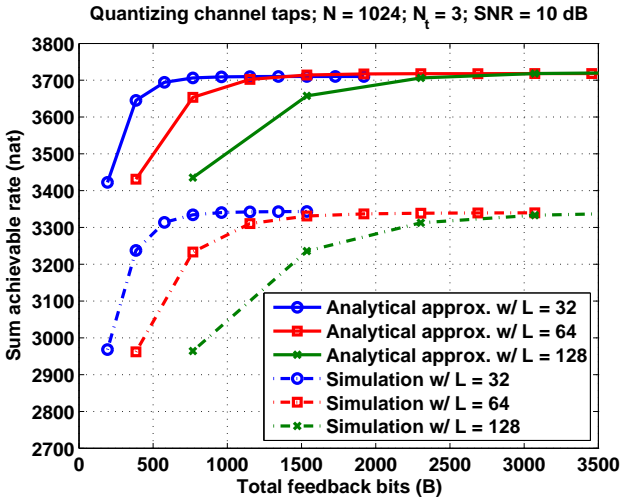}
\caption{Comparison between achievable rate obtained from simulation
  and the analytical approximation for channel-tap quantization with
  $N = 1024$, $N_t = 3$, and SNR at 10 dB.}
 \label{fig4}
\end{figure}

In Fig.~\ref{fig4}, we compare an achievable rate per subcarrier of a
$3 \times 1$ channel obtained from simulation and the
approximation~\eqref{clsim} for a direct quantization of channel taps.
A number of channel taps $L$ varies between 32 and 128. From the figure,
the approximate sum rate exhibits the same performance trend as the
simulation results and the gap between the two is about 10\%. Again we
can attribute the gap between the two results to Jensen's
inequality. Although the approximation is derived for a large feedback
rate, it seems to predict well the simulation result even with
relatively small $B$.  In addition, we observe from the simulation
results that approximately 3 bits per real coefficient are needed to
achieve close to the maximum achievable rate. While the number of
fading paths $L$ increases, $B$ also increases to achieve close to the
maximum rate.

\begin{figure}[ht]
\centering
\includegraphics[width=3.5in]{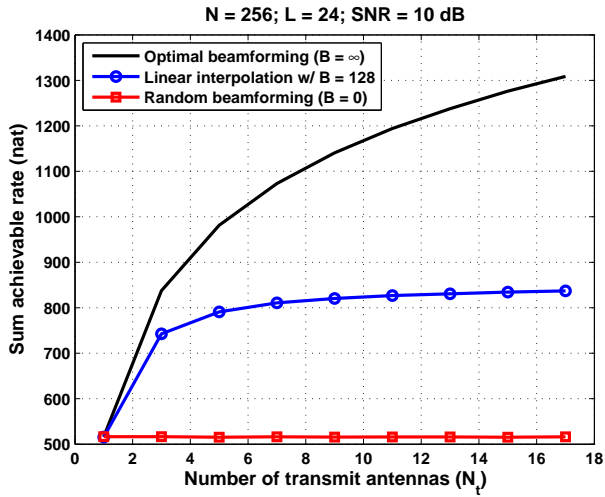}
\caption{Sum achievable rates of the proposed linear interpolation
  method, the perfect transmit beamforming, and random beamforming are
  plotted with $N_t$ for $N = 256$, $L = 24$ and SNR = 10 dB.}
 \label{f4add}
\end{figure}

\begin{figure}[ht]
\centering
\includegraphics[width=3.5in]{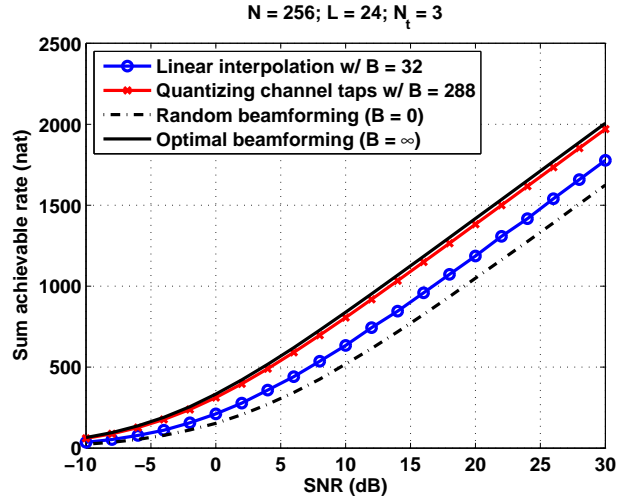}
\caption{Sum achievable rates from various methods are plotted with
SNR for $N = 256$, $L = 24$ and $N_t = 3$.}
 \label{f5add}
\end{figure}

In Figs.~\ref{f4add} and~\ref{f5add}, we compare sum rate of our
linear interpolation method with those of perfect transmit beamforming
(infinite feedback) and random transmit beamforming (zero
feedback). In Fig.~\ref{f4add}, we plot sum achievable rates with the
number of transmit antennas for $N = 256$, $L = 24$, and SNR at 10
dB. We see that the sum rates of the perfect beamforming and the
linear interpolation with $B = 128$ bits increase with $N_t$. The gap
between the perfect beamforming and our method grows larger as $N_t$
increases. To close the gap, more feedback bits are needed for
quantizing transmit beamformers.

In Fig.~\ref{f5add}, sum rates are plotted with SNR while $N_t =
3$. As expected, all sum rates increase with SNR. We also add the
performance of channel quantization with $B = 288$ or 1.125 bits per
subcarrier, which is close to that of the perfect transmit
beamforming. The linear interpolation with only 32 bits or 0.125
feedback bits per subcarrier can significantly outperform random
beamforming or a system with zero feedback.

\section{Conclusions}
\label{conclude}

We have proposed feedback methods for MISO-OFDM channels.  Beamforming
interpolation with RVQ performs well with limited feedback while
direct quantization of the channel impulse response performs well with
large feedback.  Thus, switching between the two methods for different
feedback rates is recommended. We analyzed the sum achievable rate
with constant interpolation and RVQ and showed that the analytical
results can predict the performance trend and accurately predict the
optimal cluster size. From numerical examples shown, operating at the
optimal cluster size can give a significant rate gain over an
arbitrary size. For a relatively static channel in which the number of
channel taps or SNR do not change often, the cluster size does not
have to be updated frequently as well.

For linear interpolation, we have derived a closed-form expression for
a phase rotation to avoid exhaustive search and additional number of
feedback bits in quantizing the phase rotation. We also considered the
higher-order interpolation inspired from the OFDM channel estimation
problem.  Both linear and higher-order interpolations are improved
significantly with the optimized cluster size derived for the constant
interpolation. Furthermore, we have analyzed the achievable rate with
direct quantization of channel taps, which depends on the feedback
rate and the number of antennas and channel taps.

Future work can take different directions. In the problem considered,
the MISO channel was investigated. Extending our results to MIMO
beamforming is not straightforward and thus, MIMO beamforming could be
a good problem to consider. In addition, here we considered channels
with a uniform power delay profile. Other practical channel models
might be of interest.

\appendix

\subsection{Proof of Lemma~\ref{lemma_cor}}
\label{append_lemma_cor}

We approximate
\begin{equation}
E\left[ |\bar{\bh}_n^{\dag}\bar{\bh}_{n+q}|^2 \right] \approx
\frac{E\left[
    |\bh_n^{\dag} \bh_{n+q}|^2\right]}{E\left[\|\bh_n\|^2\|\bh_{n+q}\|^2
    \right]} \label{Exsp}.
\end{equation}

First, we evaluate $E\left[ |\bh_n^{\dag} \bh_{n+q}|^2\right]$ as
follows
\begin{align}
 E&\left[|\bh_{n}^{\dag}\bh_{n+q} |^2\right]
 = E\left[ \left| \sum_{m = 1}^{N_t}h_{n,m}^*h_{n+q,m} \right|^2\right]  \\
 &= \sum_{m_1=1}^{N_t} E \left[ |h_{n, m_1}|^2|h_{n+q, m_1}|^2 \right]\nonumber\\
 &\quad+ \sum_{m_2 = 1}^{N_t}\sum_{\substack{m_2 = 1\\m_2 \neq m_1}}^{N_t}
 E[ h_{n,m_1}^*h_{n+q,m_1}] E[ h_{n,m_2}h_{n+q,m_2}^*] \\
&= N_t E\left[ |h_{n,m_1}|^2 |h_{n+q,m_1}|^2 \right] \nonumber\\
&\quad + N_t(N_t-1)|
 E\left[h_{n,m_1}^*h_{n+q,m_1} \right]|^2
 \label{a1}
\end{align}

where we apply the assumption that channel gains across antennas are
{\em i.i.d}.

Next we evaluate each term in~\eqref{a1} by substituting~\eqref{fft}.
\begin{multline}
E\left[|h_n|^2|h_{n+q}|^2 \right] \\= \sum_{l_1,l_2,l_3,l_4 = 1}^L E\left[
  g_{l_1}g_{l_2}^*g_{l_3}g_{l_4}^*\me^{\frac{-j2\pi\{(l_1-l_2+l_3-l_4)n
      +(l_3-l_4)q\}}{N}}\right]
\label{bfgdis}
\end{multline}
where we omit the antenna subscript $m_1$ for brevity. It is
straightforward to show that
\begin{multline}
 E [ g_{l_1}g_{l_2}^*g_{l_3}g_{l_4}^*
 \me^{\frac{-j2\pi\{(l_1-l_2+l_3-l_4)n +(l_3-l_4)q\}}{N}}] \\=
 \left\{ \begin{array}{l@{\quad:\quad}l} \frac{2}{L} &
   l_1=l_2=l_3=l_4,\\
   1-\frac{1}{L} & (l_1=l_2) \neq (l_3 = l_4),\\
\frac{1}{L}(\varphi^2(q)-1) & (l_1=l_3) \neq (l_2=l_4),\\
0 & \text{otherwise}.
\end{array} \right.
\label{gdis}
\end{multline}
Substituting~\eqref{gdis} into~\eqref{bfgdis} gives
\begin{equation}
 E\left[|h_n|^2|h_{n+q}|^2 \right] = 1 +
 \frac{1}{L^2}\varphi^2(q) \label{1_1}
\end{equation}

Also,
\begin{align}
E\left[h_n^*h_{n+q}\right]
&= \sum_{l_1=1}^L E|g_{l_1}|^2 \me^{\frac{-j2\pi l_1q}{N}} \nonumber\\
 &\quad +
\sum_{l_1 = 1}^L\sum_{\substack{l_2 = 1\\l_2\neq l_1}}^L
E[g_{l_1}^*g_{l_2}] \me^{\frac{-j2\pi l_1n}{N}} \me^{\frac{-j2\pi
    l_2(n+q)}{N}} \label{gg}\\
&= \sum_{l_1=1}^L \frac{1}{L} \me^{\frac{-j2\pi l_1q}{N}} \\
&= \frac{1}{L} \me^{\frac{-j2\pi (L-1)q}{N}}\varphi(q) . \label{1_2}
\end{align}
where the second term in~\eqref{gg} is equal to zero.

Substituting~\eqref{1_1} and~\eqref{1_2} into~\eqref{a1} gives
\begin{equation}
 E\left[|\bh_{n}^{\dag}\bh_{n+q}|^2\right] = N_t +
 \frac{N_t}{L^2}\varphi^2(q).
  \label{aHn}
\end{equation}
Following similar steps as the above evaluation of
$E\left[|\bh_{n}^{\dag}\bh_{n+q}|^2\right]$, we can show that
\begin{equation}
E\left[\|\bh_n\|^2\|\bh_{n+q}\|^2 \right] = N_t^2 + \frac{N_t}{L^2}
\varphi^2(q). \label{aHnq}
\end{equation}

Substituting~\eqref{aHn} and~\eqref{aHnq} into~\eqref{Exsp} yields the
Lemma.

\subsection{Proof of Proposition~\ref{t1}}
\label{append_prop}

From~\eqref{cap_rvq},
\begin{align}
 C_{n+q} &= E \left[ \log ( 1 + \rho \|\bh_{n+q}\|^2|\bhb_{n+q}^{\dag}
 \bvh_{n+q}|^2 ) \right] \\
 &\leq  \log ( 1 + \rho E[\|\bh_{n+q}\|^2|\bhb_{n+q}^{\dag}
 \bvh_{n+q}|^2] ) \label{leql} \\
 &= \log ( 1 + \rho E\|\bh_{n+q}\|^2 E|\bhb_{n+q}^{\dag}\bvh_{n+q}|^2 ) \label{l1E}\\
 &= \log ( 1 + \rho N_t E|\bhb_{n+q}^{\dag}\bvh_{n+q}|^2 ) \label{cap_rvq1}
\end{align}
where Jensen's inequality is applied in~\eqref{leql}. Eq~\eqref{l1E}
is due to the fact that $\|\bh_{n+q}\|^2$ and
$|\bhb_{n+q}^{\dag}\bvh_{n+q}|^2$ are
independent~\cite{yeung_miso}. In addition, $E \|\bh_{n+q}\|^2 = N_t$
since each element in $\bh_{n+q}$ has unit variance. Jensen's
inequality is tighter when the number of transmit antennas increases.

To derive the upper bound on $C_{n+q}$ in~\eqref{cap_rvq1}, we need to
determine $E|\bhb_{n+q}^{\dag} \bvh_{n+q}|^2$. With constant
interpolation, $\bvh_{n+q}$ is set to equal the representative
beamforming of a cluster, which is $q$ subcarriers away. Therefore, we
would like to evaluate $E|\bhb_{n+q}^{\dag} \bvh_{n}|^2$. To
accomplish this goal, we project $\bhb_{n+q}$ onto $\bhb_{n}$ and its
$N_t -1$-dimensional orthogonal complement denoted by $\bhb_{n}^\perp$.

Let $\{\bu_1, \bu_2, \ldots, \bu_{N_t-1}\}$ be a basis of
$\bhb_{n}^\perp$. Hence, we can write $\bhb_{n+q}$ as a linear
combination of its projection onto $\bhb_{n}$ and the basis of
$\bhb_{n}^\perp$ as follows.
\begin{equation}
\bhb_{n+q} = (\bhb_{n}^{\dag} \bhb_{n+q})\bhb_{n} + \sum_{i=1}^{N_t -
  1}(\bu_i^\dag \bhb_{n+q}) \bu_i . \label{buu}
\end{equation}
With~\eqref{buu}, we have
\begin{align}
E &\left[|\bhb_{n+q}^{\dag}\bvh_n|^2\right] \nonumber\\
&= E\left| (\bhb_{n+q}^\dag \bhb_{n})(\bhb_{n}^{\dag}\bvh_n) +
\sum_{i=1}^{N_t - 1}(\bhb_{n+q}^\dag \bu_i)(\bu_i^{\dag}\bvh_n)\right|^2 \\
&= E\left[ |\bhb_{n+q}^\dag\bhb_{n}|^2 |\bhb_{n}^{\dag}\bvh_n|^2
\right] + \sum_{i=1}^{N_t - 1} E\left[|\bhb_{n+q}^{\dag}\bu_i|^2
|\bu_i^{\dag}\bvh_n|^2  \right] \nonumber \\
&\quad + 2 E \Re\left\{ (\bhb_{n+q}^\dag \bhb_n)
(\bhb_n^{\dag}\bvh_n)\sum_{i=1}^{N_t-1}(\bhb_{n+q}^{\dag}\bu_i)
(\bu_i^{\dag}\bvh_n) \right\} \label{3Terms}
\end{align}
where $\Re\left\{ x \right\}$ is the real part of $x$. Similar
to~\cite{yeung_miso}, it can be shown that
$|\bhb_{n+q}^{\dag}\bhb_{n}|^2$ and $|\bhb_{n}^{\dag}\bvh_n|^2$ are
independent. In~\cite{yeung_miso}, $E|\bhb_{n}^{\dag}\bvh_n|^2$ was
also analyzed while $E|\bhb_{n+q}^{\dag}\bhb_n|^2 \approx \psi(q,N_t)$
from Lemma~\ref{lemma_cor}. Thus,
\begin{align}
 E[ |\bhb_{n+q}^{\dag}\bhb_{n}|^2
 &|\bhb_{n}^{\dag}\bvh_n|^2] \nonumber\\
 &=  E|\bhb_{n+q}^{\dag}\bhb_{n}|^2
 E|\bhb_{n}^{\dag}\bvh_n|^2 \\
 &\approx \psi(q,N_t) \left( 1 - 2^{B/K}
\beta(2^{B/K},\frac{N_t}{N_t-1}) \right) .\label{term1}
\end{align}

For the second term on the right-hand side of~\eqref{3Terms}, we have
that similar to the first term,
\begin{equation}
  E\left[|\bhb_{n+q}^{\dag}\bu_i|^2 |\bu_i^{\dag}\bvh_n|^2 \right] =
  E|\bhb_{n+q}^{\dag}\bu_i|^2 E|\bu_i^{\dag}\bvh_n|^2 .
\label{ElE}
\end{equation}

We can evaluate the second term in~\eqref{ElE} as follows.  Similar
to~\eqref{buu}, we can write $\bvh_n$ as a linear combination of its
projection onto basis $\{\bhb_n, \bu_1, \ldots, \bu_{N_t-1}\}$ as
follows:
\begin{equation}
  \bvh_n = (\bhb_n^\dag \bvh_n) \bhb_n + \sum_{i = 1}^{N_t -1}
  (\bu_i^\dag \bvh_n) \bu_i . \label{bvhb}
\end{equation}
Evaluating $(\bvh_n^\dag \bvh_n)^2$ with~\eqref{bvhb} and applying the
fact that $\| \bvh_n \| = 1$ results in
\begin{equation}
  |\bhb_n^{\dag}\bvh_n|^2 + \sum_{i=1}^{Nt-1} |\bu_i^{\dag}\bvh_n|^2 = 1.
\label{Sph}
\end{equation}
We take expectation on both sides and substitute a closed-form
expression of $E|\bhb_n^{\dag} \bvh_n|^2$
from~\cite{yeung_miso}. Also, $E|\bu_i^\dag \bvh_n|^2$ is the same for
all $1 \le i \le N_t - 1$ due to identical distributions.  Thus,
from~\eqref{Sph}, we have
\begin{equation}
E |\bu_i^{\dag}\bvh_n|^2 = \frac{2^{B/K}
\beta(2^{B/K},\frac{N_t}{N_t-1})}{N_t-1} \label{ExPer}.
\end{equation}

Similar to the steps that derive~\eqref{ExPer}, we can show that
\begin{equation}
  E|\bhb_{n+q}^{\dag}\bu_i|^2 \approx \frac{1 - \psi(q,N_t)}{N_t - 1}
  . \label{bhbpsi}
\end{equation}
Applying~\eqref{ExPer} and~\eqref{bhbpsi}, we have

\begin{multline}
\sum_{i=1}^{N_t-1} E\left[ |\bhb_{n+q}^{\dag}\bu_i|^2
  |\bu_i^{\dag}\bvh_n|^2 \right] \\
\approx (1-\psi(q,N_t))\cdot
\frac{2^{B/K} \beta(2^{B/K},\frac{N_t}{N_t-1})}{N_t-1}
\label{term2}
\end{multline}

Evaluating the final term of the right-hand side of~\eqref{3Terms} is
not tractable. However we note that for both small and large feedback,
the term is close to zero due to $\bhb_n^{\dag}\bvh_{n}$ and
$\bu_i^{\dag}\bvh_{n}$, respectively. Thus, we approximate
\begin{equation}
E \Re\left\{ (\bhb_{n+q}^\dag \bhb_n) (\bhb_n^{\dag}\bvh_n)
\sum_{i=1}^{N_t-1}(\bhb_{n+q}^{\dag}\bu_i) (\bu_i^{\dag}\bvh_n)
\right\} \approx 0 . \label{Ere}
\end{equation}

Finally, substituting~\eqref{term1}, \eqref{term2}, and~\eqref{Ere}
in~\eqref{3Terms} yields Proposition~\ref{t1}.

\subsection{Proof of Proposition~\ref{c2}}
\label{pp2}

Applying the linear interpolation~\eqref{intv} and assuming optimal,
{\em unquantized} beamforming, we have
\begin{multline}
E|(\bv_{kM}^\opt)^\dag \bv_{kM + m}|^2 \\ \approx
 \frac{E\left|\bh_{kM}^{\dag}\left\{(1-c_m) \bh_{kM}+c_m \me^{j\theta_m}
    \bh_{(k+1)M}\right\}\right|^2}{E\|(1-c_m)\bh_{kM}+ c_m
    \me^{j\theta_m}\bh_{(k+1)M}\|}.
    \label{Esp}
\end{multline}

Here we propose to set phase rotation $\theta_m$ by solving
\begin{equation}
   \frac{E\left|\bh_{kM}^{\dag}\left\{(1-c_m) \bh_{kM}+c_m
       \me^{j\theta_m}
       \bh_{(k+1)M}\right\}\right|^2}{E\|(1-c_m)\bh_{kM}+
       c_m \me^{j\theta_m}\bh_{(k+1)M}\|} = \psi(m, N_t) .
\end{equation}
where $\psi(m, N_t)$ is defined in Lemma~\ref{lemma_cor}.

Similar to steps shown in the proof of Lemma~\ref{lemma_cor}, we can
show that
\begin{multline}
E\left|\bh_{kM}^{\dag}\left\{(1-c_m) \bh_{kM}+c_m
\me^{j\theta_m} \bh_{(k+1)M}\right\}\right|^2 \\ =
(1-c_m)^2(N_t+1) + c_m^2\left( \frac{N_t}{L} \varphi(M)^2 + 1
\right)\\ + 2(1-c_m)c_m\cos\theta_m\cos\left( \frac{\pi M(L-1)}{N}
\right)\left(\frac{N_t + 1}{L} \right) \varphi(M) \label{Eb}
\end{multline}
and
\begin{multline}
E\|(1-c_m)\bh_{kM}+ c_m \me^{j\theta_m}\bh_{(k+1)M}\| \\=
(1-c_m)^2(N_t + 1) + c_m^2N_t \\+
2\frac{N_t}{L}(1-c_m)c_m\cos\theta_m\cos\left( \frac{\pi M(L-1)}{N}
\right) \varphi(M) \label{El}.
\end{multline}
Substituting~\eqref{Eb} and~\eqref{El} into~\eqref{Esp} and solving
for $\theta_m$ gives Proposition~\ref{c2}.

\bibliographystyle{myIEEEtran}
\bibliography{IEEEabrv,MisoOfdm}

\begin{thebibliography}{10}
\providecommand{\url}[1]{#1}
\csname url@samestyle\endcsname
\providecommand{\newblock}{\relax}
\providecommand{\bibinfo}[2]{#2}
\providecommand{\BIBentrySTDinterwordspacing}{\spaceskip=0pt\relax}
\providecommand{\BIBentryALTinterwordstretchfactor}{4}
\providecommand{\BIBentryALTinterwordspacing}{\spaceskip=\fontdimen2\font plus
\BIBentryALTinterwordstretchfactor\fontdimen3\font minus
  \fontdimen4\font\relax}
\providecommand{\BIBforeignlanguage}[2]{{%
\expandafter\ifx\csname l@#1\endcsname\relax
\typeout{** WARNING: IEEEtran.bst: No hyphenation pattern has been}%
\typeout{** loaded for the language `#1'. Using the pattern for}%
\typeout{** the default language instead.}%
\else
\language=\csname l@#1\endcsname
\fi
#2}}
\providecommand{\BIBdecl}{\relax}
\BIBdecl

\bibitem{lo99}
T.~K.~Y. Lo, ``Maximum ratio transmission,'' \emph{{IEEE} Trans. Commun.},
  vol.~47, no.~10, pp. 1458--1461, Oct. 1999.

\bibitem{mimo}
W.~Santipach and M.~L. Honig, ``Capacity of a multiple-antenna fading channel
  with a quantized precoding matrix,'' \emph{{IEEE} Trans. Inf. Theory},
  vol.~55, no.~3, pp. 1218--1234, Mar. 2009.

\bibitem{love06}
D.~J. Love, R.~W. Heath, Jr., V.~K.~N. Lau, D.~Gesbert, B.~D. Rao, and
  M.~Andrews, ``An overview of limited feedback in wireless communication
  systems,'' \emph{{IEEE} J. Sel. Areas Commun.}, vol.~26, no.~8, pp.
  1341--1365, Oct. 2008.

\bibitem{ClusterWu}
M.~Wu, C.~Zhen, and Z.~Qui, ``Feedback reduction based on clustering in
  {MIMO-OFDM} beamforming systems,'' in \emph{Proc. IEEE Int. Conf. on Wireless
  Commun., Networking and Mobile Computing (WiCom)}, Beijing, China, Sep. 2009,
  pp. 1--4.

\bibitem{choi}
J.~Choi and R.~W. Heath, Jr., ``Interpolation based transmit beamforming for
  {MIMO-OFDM} with limited feedback,'' \emph{{IEEE} Trans. Signal Process.},
  vol.~53, no.~11, pp. 4125 -- 4135, Nov. 2005.

\bibitem{he11}
C.~He, P.~Zhu, B.~Sheng, and X.~You, ``Two novel interpolation algorithms for
  {MIMO-OFDM} systems with limited feedback,'' in \emph{Proc. IEEE Veh.
  Technol. Conf. (VTC Fall)}, San Francisco, CA, USA, Sep. 2011. ISSN 1090-3038
  pp. 1--5.

\bibitem{huang11}
Q.~Huang, M.~Ghogho, Y.~Li, D.~Ma, and J.~Wei, ``Transmit beamforming for
  {MISO} frequency-selective channels with per-antenna power constraint and
  limited-rate feedback,'' \emph{{IEEE} Trans. Veh. Technol.}, vol.~60, no.~8,
  pp. 3726--3735, Oct. 2011.

\bibitem{ye}
H.~Z. Ye, V.~Stolpaman, and N.~van Waes, ``A reduced {CSI} feedback approach
  for precoded {MIMO-OFDM} system,'' \emph{{IEEE} Trans. Wireless Commun.},
  vol.~6, no.~1, pp. 55--58, Jan. 2007.

\bibitem{long12}
H.~Long, K.~J. Kim, W.~Xiang, S.~Shen, K.~Zheng, and W.~Wang, ``Improved
  wideband precoding with arbitrary subcarrier grouping in {MIMO-OFDM}
  systems,'' \emph{IET Commun.}, vol.~6, no.~3, pp. 281--288, Feb. 2012.

\bibitem{pande}
T.~Pande, D.~J. Love, and J.~V. Krogmeier, ``Reduced feedback {MIMO-OFDM}
  precoding and antenna selection,'' \emph{{IEEE} Trans. Signal Process.},
  vol.~55, no.~5, pp. 2284--2293, May 2007.

\bibitem{huang08}
J.~Huang, J.~Zhang, Z.~Liu, J.~Li, and X.~Li, ``Transmit beamforming for
  {MIMO-OFDM} systems with limited feedback,'' in \emph{Proc. IEEE Veh.
  Technol. Conf. (VTC Fall)}, Calgary, BC, Sep. 2008, pp. 1 -- 5.

\bibitem{kim10}
J.~Kim, B.~Kim, J.~Lee, and D.~Park, ``Closed-form interpolations for
  {MISO-OFDM} beamforming codewords,'' in \emph{Proc. Int. Conf. on Inf. and
  Commun. Technol. Convergence (ICTC)}, Jeju, Korea, Nov. 2010, pp. 43--44.

\bibitem{he12}
H.~He and Y.~Zeng, ``Transmit beamforming interpolation algorithm for
  {MIMO-OFDM} systems in the limited feedback scenario,'' in \emph{Proc. IEEE
  Int. Conf. on Computer and Inf. Technol. (CIT)}, Chengdu, Sichuan, China,
  Oct. 2012, pp. 696 -- 699.

\bibitem{chang12}
J.~Chang, I.~T. Lu, and Y.~X. Li, ``Adaptive codebook-based channel prediction
  and interpolation for multiuser multiple-input multiple-output orthogonal
  frequency division multiplexing systems,'' \emph{ETRI Journal}, vol.~34,
  no.~1, pp. 9--16, Feb. 2012.

\bibitem{wang12}
L.~Wang and C.~H. B.~Sheng, ``Study on feedback reduction techniques in
  {MIMO-OFDM} beamforming systems,'' in \emph{Proc. Int. Conf. on Wireless
  Commun. and Signal Process. (WCSP)}, Huangshan, Anhui, China, Oct. 2012, pp.
  1--5.

\bibitem{zhou06}
S.~Zhou, B.~Li, and P.~Willett, ``Recursive and trellis-based feedback
  reduction for {MIMO-OFDM} with rate-limited feedback,'' \emph{{IEEE} Trans.
  Wireless Commun.}, vol.~5, no.~12, pp. 3400--3405, Dec. 2006.

\bibitem{zhao12}
B.~Zhao, T.~Akbudak, M.~Simsek, A.~Czylwik, and H.~Xu, ``Limited feedback for
  {MISO-OFDM} systems,'' in \emph{Proc. Int. OFDM Workshop (InOWo)}, Duisburg,
  Germany, Aug. 2012, pp. 1--4.

\bibitem{ghirmai14}
T.~Ghirmai, ``Design of reduced complexity feedback precoding for
  {MIMO-OFDM},'' \emph{Int. J. Commun. Syst.}, 2014. doi: 10.1002/dac.2806

\bibitem{liu92}
G.~S. Liu and C.~H. Wei, ``A new variable fractional sample delay filter with
  nonlinear interpolation,'' \emph{{IEEE} Trans. Circuits Syst. {II}}, vol.~93,
  no.~2, pp. 1057--7130, Feb. 1992.

\bibitem{hsieh98}
M.~H. Hsieh and C.~H. We, ``Channel estimation for {OFDM} systems based on
  comb-type pilot arrangement in frequency selective fading channels,''
  \emph{{IEEE} Trans. Consum. Electron.}, vol.~44, no.~1, pp. 217 -- 225, Feb.
  1998.

\bibitem{he09}
C.~He, Z.~Peng, Q.~Zeng, and Y.~Zeng, ``A novel {OFDM} interpolation algorithm
  based on comb-type pilot,'' in \emph{Proc. IEEE Int. Conf. on Wireless
  Commun., Networking and Mobile Computing (WiCom)}, Beijing, China, Sep. 2009,
  pp. 893--896.

\bibitem{diallo13}
M.~Diallo, M.~H\'{e}lard, and L.~Cariou, ``A limited and efficient quantized
  feedback for {IEEE} 802.11n evolution,'' in \emph{Proc. IEEE Int. Conf. on
  Telecommun. (ICT)}, Casablanca, Morocco, May 2013, pp. 1--5.

\bibitem{ecti12}
K.~Mamat and W.~Santipach, ``Subcarrier clustering for {MISO-OFDM} channels
  with quantized beamforming,'' in \emph{Proc. Int. Conf. on Electrical
  Engineering/Electronics, Computer, Telecommun. and Inf. Technol. (ECTI-CON)},
  Huahin, Thailand, May 2012, pp. 1--4.

\bibitem{icc13}
------, ``On transmit beamforming for multiantenna {OFDM} channels with
  finite-rate feedback,'' in \emph{Proc. IEEE Int. Conf. on Commun. (ICC)},
  Budapest, Hungary, Jun. 2013, pp. 5689--5693.

\bibitem{yeung_miso}
C.~K. Au-Yeung and D.~J. Love, ``On the performance of random vector
  quantization limited feedback beamforming in a {MISO} system,'' \emph{{IEEE}
  Trans. Wireless Commun.}, vol.~6, no.~2, pp. 458--462, Feb. 2007.

\bibitem{borah99}
D.~K. Borah and B.~D. Hart, ``Frequency-selective fading channel estimation
  with a polynomial time-varying channel model,'' \emph{{IEEE} Trans. Commun.},
  vol.~47, no.~6, pp. 862--873, Jun. 1999.

\bibitem{nguyen07}
V.~Nguyen, H.-P. Kuchenbecker, H.~Haas, K.~Kyamakya, and G.~Gelle, ``Channel
  impulse response length and noise variance estimation for {OFDM} systems with
  adaptive guard interval,'' \emph{{EURASIP} J. on Wireless Commun. and
  Networking}, 2007. doi: 10.1155/2007/24342

\bibitem{commag04}
D.~J. Love, R.~W. Heath, Jr., W.~Santipach, and M.~L. Honig, ``What is the
  value of limited feedback for {MIMO} channels?'' \emph{{IEEE} Commun. Mag.},
  vol.~42, no.~10, pp. 54--59, Oct. 2004.

\bibitem{jayant84}
N.~S. Jayant and P.~Noll, \emph{Digital Coding of Waveforms: {P}rinciples and
  Applications to Speech and Video}.\hskip 1em plus 0.5em minus 0.4em\relax
  Englewood Cliffs, NJ, USA: Prentice-Hall, 1984.

\bibitem{zhang14}
Q.~Zhang, S.~Jin, K.-K. Wong, H.~Zhu, and M.~Matthaiou, ``Power scaling of
  uplink massive {MIMO} systems with arbitrary-rank channel means,''
  \emph{{IEEE} J. Sel. Topics Signal Process.}, vol.~8, no.~55, pp. 966--981,
  Oct. 2014.

\bibitem{hui01}
D.~Hui and D.~L. Neuhoff, ``Asymptotic analysis of optimal fixed-rate uniform
  scalar quantization,'' \emph{{IEEE} Trans. Inf. Theory}, vol.~47, no.~3, pp.
  957--977, Mar. 2001.

\bibitem{bucklew79}
J.~Bucklew and N.~Gallagher, Jr., ``A note on optimal quantization,''
  \emph{{IEEE} Trans. Inf. Theory}, vol.~25, no.~3, pp. 365--366, May 1979.

\end{thebibliography}

\end{document}